\pdfoutput=1
\documentclass[a4paper,referee]{raa}            % referee version: for submission
\usepackage{multirow}
\usepackage{graphicx,times}             %for PS/EPS graphics inclusion, new
\usepackage{natbib}
\usepackage{amssymb,amsmath}

\bibpunct{}{}{;}{a}{}{,}

\usepackage[pagebackref=true]{hyperref}

\usepackage{booktabs}

\begin{document}

\title{Searching for kilonova with the SiTian prototype telescope}
    \volnopage{Vol.0 (20xx) No.0, 000--000}      %%preserved for Editor. DOn't remove!
    \setcounter{page}{1}          %%starting page, preserved for Editor. DOn't remove!
    \author{Zhirui Li
        \inst{1,2}
        , Hongrui Gu\inst{1,2}
        ,
        Yongkang Sun\inst{1,2},
        Yang Huang\inst{2,1,4}
        , Youjun Lu\inst{2,3}
        $\And$
        Jifeng Liu\inst{1,2,4}
    }
    \institute{
    \item{\inst{1}New Cornerstone Science Laboratory, National Astronomical Observatories CAS, Beijing 100101, People's Republic of China}\\
    \item{\inst{2}School of Astronomy and Space Sciences, University of Chinese Academy of Sciences, Beijing 100158, People's Republic of China}
    \item{\inst{3}National Astronomical Observatories, Chinese Academy of Sciences, Beijing 100101, People's Republic of China}
    \item{\inst{4}Corresponding authors: huangyang@ucas.ac.cn; jfliu@nao.cas.cn}
             }
\date{May 2023}

\abstract{\\
We simulate the optical searching of gravitational-wave electromagnetic counterpart of the binary neutron star (BNS) merger event (i.e., a kilonova) using the ground based {\it SiTian} project prototype telescope with a 5-min limiting magnitude of 22.0, triggered by LIGO and Virgo gravitational wave detectors during the ongoing O4 run. Our simulations show that an average of 0.17-0.25 kilonova events can be observed over the entire O4 period of $\sim 2$ years in the most optimistic case we set, while no kilonova can be detected in other cases. We note that it is beneficial for {\it SiTian}'s kilonova searching by extending the exposure time to gain deeper limiting magnitude despite the rapid decline of kilonova luminosity.}

\maketitle

\section{Introduction}

The pioneering detection of the first gravitational wave (GW) associated kilonova event was achieved during the LIGO-Virgo (LV) Observation Run 2 (O2), marking a milestone in the field of multi-messenger gravitational wave astronomy (\cite{Abbott2017}). LIGO (\cite{Abbott_2009}) successfully detected the GW event, GW170817, and approximately 2 seconds later the gamma-ray burst (GRB) event, GRB170817A, triggered the GRB monitor on the Fermi satellite (\cite{Page2020}). Subsequently, the gravitational wave electromagnetic counterpart (GWEM) known as AT2017gfo was observed in the host galaxy NGC 4993, $\sim 11$ hours later by Las Cumbres Observatory (LCO) 1-meter telescopes (e.g., \cite{Arcavi2017}, \cite{Soares2017}, \cite{Tanvir2017}, \cite{Cowperthwaite2017}). This GWEM transient, referred to as a kilonova, is believed to be associated with the $r$-process occurring in neutron-rich ejecta generated from a binary neutron star (BNS) merger. A kilonova is theoretically the brightest optical event among GWEMs under most circumstances (e.g., \cite{Meyer1989}, \cite{Davies1994}, \cite{Li1998}, \cite{Rosswog2005}, \cite{Metzger2010}, \cite{Kasen2013}), which allows kilonova to provide valuable insights into the origin and evolution of neutron stars, their mass distribution and equation of states, investigating the enrichment of $r$-process elements in their host galaxies (\cite{Metzger2010}) and may serve as standard sirens for measuring cosmological parameters within the local universe (e.g., \cite{Abbott2017}, \cite{Coughlin_2020}).

Limited by the detection method of the current ground-based GW detectors, the spatial positions of the confirmed GW events are quite uncertain, spanning dozens to over several thousand square degrees. As a result, the ultraviolet/optical/infrared (hereafter UV/O/IR) time-domain survey has played a crucial role in facilitating follow-up observations of GWEM. Numerous UV/O/IR observing projects, including ATLAS (\cite{Tonry2018}), GRANDMA (\cite{Agayeva_2021}), DES-GW (\cite{Herner_2017}), Wide-Field Survey Telescope (WFST, \cite{Wang_2023}), Zwicky Transient Facility (ZTF, \cite{Bellm_2018}), All-Sky Automated Survey for Supernovae (ASAS-SN, \cite{Kochanek2017}), Ground Wide-Angle Camera network (GWAC, \cite{Han_2021}) etc., have undertaken GWEM follow-up efforts for LVK-O2, O3 and ongoing O4 observations. However, apart from this groundbreaking discovery of GW170817, no more GWEMs have been detected during the left phase of LIGO-Virgo O2, the entire LIGO-Virgo O3 observation runs, and the ongoing LIGO-Virgo-KAGRA (LVK) O4 runs. As a result, AT2017gfo remains the sole kilonova event with a recorded GW signal to date.

The luminosity of a kilonova in the optical band can be significantly diminished by the presence of neutron-rich ejecta, which introduces a high level of uncertainty and potential anisotropy on ejecta distributions (\cite{Metzger2010}, \cite{Kasen2013}, \cite{Bulla2019}). Consequently, the maximum luminosity of a kilonova may experience substantial attenuation, causing a significant fraction of kilonova to fall below the detection limit of most UV/O/IR wide-field survey telescopes, particularly for those with smaller apertures such as the ASAS-SN and GWAC. Moreover, conventional optical transient surveys carried out with large apertures frequently encounter the limited field of view (FoV). Therefore, multiple exposures may be required to accurately pinpoint the kilonova's location, led to observations obtained when its luminosity has significantly faded and becomes unobservable. Therefore, the existing optical time-domain surveys have faced considerable constraints in accurately monitoring rapidly changing transient phenomena like kilonova, which constitute major barriers to the identification and investigation of kilonova using current UV/O/IR surveys.

To address these challenges, the {\it SiTian} project (\cite{Jifeng2021}) proposes an innovative approach. It involves the establishment of several ground-based nodes, each housing 20 groups of 60 wide-field optical survey telescopes. Each group comprises three telescopes equipped with SDSS-like {\it gri} filters. These telescopes are 1-m f/2 wide-field refractors, equipped with a science-grade CMOS camera at the back end, utilizing a $2\times2$ stitched configuration. Consequently, each telescope offers a sizable FoV spanning 25 square degrees at $5^\circ\times5^\circ$ and exhibits a deep survey capability, reaching beyond the magnitude of 21 within a 1-minute exposure. Such a design holds promise for significantly improving the search efficiency and sensitivity to kilonova.

The {\it SiTian} project focuses on various aspects of transient sources, short-timescale variable sources, stellar activity, and other frontier studies related to short-timescale time-domain surveys. With its large FoV, deep observational capabilities, and high survey efficiency, the {\it SiTian} project has the potential to bridge the observational gap between the shallow limiting magnitude of wide angle small aperture surveys and low efficiency of large aperture telescope surveys in kilonova studies, and play a crucial role in future kilonova searching surveys and other GWEM investigations.

Currently, the first {\it SiTian} prototype telescope, installed with an SDSS $g$-band filter, has been deployed at the Xinglong Observatory and already had the first-light in late 2024. Furthermore, the first multi-color group, comprising three prototype telescopes, is planned to be installed at the Lenghu Observatory in early 2025. The development of these instruments are anticipated to significantly contribute to GWEM follow-up observations during the upcoming LVK-O4+ run. Meanwhile, the ongoing LVK O4 is expected to upgrade the detection capabilities for BNS mergers to a distance of 160-190 Mpc for single LIGO Hanford / Livingston GW detector at a GW signal to noise ratio of 8 ({\tt https://dcc-lho.ligo.org/LIGO-T2000012/public}). It is indeed a right time to search GWEM events through the first prototype yet powerful single telescope of {\it SiTian} project. 

In this paper, we present preliminary results on the capacity to GWEM follow-up observations of the single {\it SiTian} prototype telescope installed at the Xinglong Observatory. These results are based on primordial data simulations utilizing the designed parameters of the {\it SiTian} prototype telescopes, as well as the historical weather conditions of Xinglong site (e.g., meteorological data, limiting magnitude under different lunar phase and distance, etc.). Section 2 provides a detailed discussion of our methodology for the generation and simulation of mock BNS systems, including the spatial distribution of BNSs, simulated ground-based GW observations of BNS mergers, and the light curves of their counterparts, etc. Section 3 discusses our dataset of mock LVK-O4 BNS mergers with corresponding light curves obtained from simulations, as well as the observation analysis utilizing single {\it SiTian} telescope. Section 4 presents a summary and suggestions for further simulations and observations with the {\it SiTian} project.

\section{Methods and presets of Simulation}

In this study, we conduct a simulation of {\it SiTian}'s GWEM search process in two main steps. First, we generate multi-messenger signals for BNS mergers and kilonova. Second, we simulate {\it SiTian}'s planning and execution of kilonova searches at the Xinglong Observatory using mock data generated by the first step.

Given the notably low event rate of BNS mergers, simulating a single O4 observing period is statistically inadequate. To address this, we distribute them across various O4 simulation subsets and calculate an average based on the given event rate. This method is built on the premise that the occurrence rate of BNS mergers is independent of the BNS properties, allowing us to model a large set number of BNS merger cases, distribute them across various O4 simulation subsets, and calculate an average based on the given event rate, resulting in a Monte Carlo-based result.

This section discusses the generation of BNS merger events and their multi-messenger signal observations.

\subsection{Parameters for generating BNS merger events}

It is recognized that BNS mergers emit a series of multi-messenger signals. In this study, we consider only the two most important of them, namely the high-frequency gravitational wave (GW) signals, and the optical SDSS-{\it g} band kilonova lightcurve that {\it SiTian} would acquired when simulating the conduction of the search efforts.

To obtain these two signals for an arbitrary BNS merger event, we used the Python package {\tt bilby} (\cite{Ashton2019}) to generate gravitational waveforms with ground-based simulated observations, and the Python package {\tt gwemlightcurves} (\cite{Coughlin2018}) to generate the kilonova lightcurves. With certain assumptions, including the direction of rotation of the two neutron stars (NSs) in the BNS system is consistent with the orbital rotation axis, and the lightcurves of kilonovas are modeled using the {\tt Me2017} model with the velocity exponent of the ejecta $\beta=3$, where $\beta$ defined as $M_\mathrm{ejecta}=v_\mathrm{ejecta}^\beta$, and $M_\mathrm{ejecta}$ is the ejecta mass and $v_\mathrm{ejecta}$ the ejecta velocity. The parameters that we require for the BNS merger generation are summarized in Table 1.

It can be seen from Table 1 that in order to generate a complete multi-messenger signal for a BNS merger event, a total of 19 parameters are in need to be generated. Still, many of the parameters can be chosen randomly, which we have marked in Table 1. Furthermore, many of the parameters are non-independent, e.g., the compactness $C$ of a NS is closely related to its gravitational mass. Therefore, the number of parameters that we need to generate specifically is reduced to 7, respectively the redshift $z$, the Right Ascension and Declination of the BNS system, the time $t$ at which the BNS merger occurs, the kilonova intrinsic opacity $\kappa$, and the gravitational mass of the companion star $m_{g1}$, $m_{g2}$ in the BNS system. We discuss the generation of these parameters in the following text.

\begin{table}
\centering
\caption{Parameters used for generating BNS merger events.}
\label{tab:BNS_parameters}
\begin{tabular}{@{}lccc@{}}
\toprule
\textbf{Quantity used in our simulation} & \textbf{GW signal} & \textbf{Kilonova signal} & \textbf{Uniformly generated} \\ \midrule
\multicolumn{1}{l|}{\textit{\textbf{Binary properties}}} &  &  &  \\
\multicolumn{1}{l|}{Right Ascension (R.A.)} & Y & Y & - \\
\multicolumn{1}{l|}{Declination (Dec.)} & Y & Y & - \\
\multicolumn{1}{l|}{Redshift ($z$)} & Y & Y & - \\
\multicolumn{1}{l|}{Time of merger ($t$)} & Y & Y & Y* \\
\multicolumn{1}{l|}{Polarisation angle ($\psi$)} & Y & - & Y \\
\multicolumn{1}{l|}{$\phi_0$} & Y & - & Y \\
\multicolumn{1}{l|}{$\theta_{jn}$} & Y & Y & Y \\
\multicolumn{1}{l|}{Coalescence phase ({\tt coa\_phase})} & Y & - & Y \\
\multicolumn{1}{l|}{Opacity ($\kappa$)} & - & Y & - \\ \midrule
\multicolumn{1}{l|}{\textit{\textbf{NS properties}}} &  &  &  \\
\multicolumn{1}{l|}{Gravitational mass ($M_g$)} & Y & Y & - \\
\multicolumn{1}{l|}{Baryonic mass ($M_b$)} & - & Y & - \\
\multicolumn{1}{l|}{Dimensionless spin ($a$)} & Y & - & Y \\
\multicolumn{1}{l|}{Compactness ($C$)} & - & Y & - \\
\multicolumn{1}{l|}{Tidal deformability ($\Lambda_D$)} & Y & Y & - \\ \bottomrule
\multicolumn{1}{l}{* Time is generated with rules discussed in the text.}
\end{tabular}
\end{table}

We begin by discussing the redshift $z$. Suppose that the approximate distribution $p(z)$ of BNS merger rates with respect to $z$ satisfies

$$p(z)=\mu\frac{\mathrm{d}V_c}{\mathrm{d}z}\frac{1}{1+z};$$
where $V_c$ is the co-moving volume, in which we adopt the cosmological parameters recommended by Planck 2018 (\cite{Planck2020}) in our calculations; and $\mu$ the localized BNS merger event rate. It is easy to find that $p(z)$ is not convergent, so we need to designate an upper bound for the distribution of the BNS. In this simulation, the upper bound is taken to be 500 Mpc, which corresponds approximately to a redshift $z = 0.1$.

When generating the right ascension and declination for BNS mergers, we assume they occur within galaxies. Given that cosmological isotropy is not obvious within a distance of 500 Mpc, the common supposition of an isotropic spatial distribution for BNS mergers is deemed unrealistic. Thus, we derive their positions by linking the redshift $z$ of each BNS system to actual galaxies in a catalog. Specifically, we employ the {\tt GLADE+} galaxy catalog (\cite{D_lya_2022}), which is $>60\%$ complete for galaxies within 500 Mpc across much of the sky, except for a selection bias near the Galactic plane within 20 degrees due to Galactic extinction effects. Despite this bias, we consider the use of this catalog sufficient to yield credible simulation results.

Regarding the time of merger events $t$, it is reasonable to suppose that these occur uniformly throughout the observation period. In this simulation, we uniformly and randomly distribute each BNS merger event over the duration between the start of the O4a observations at $\mathrm{MJD}=60087$ and the end of the O4b observations at $\mathrm{MJD}=60835$. Based on an assumed co-moving volume rate of 100 BNS merger events per year within $1\ \mathrm{Gpc}^3$, this translates to roughly 32 BNS merger events annually within 500 Mpc. However, accounting for the {\tt GLADE+} catalog's selection bias against regions near the Galactic plane, we estimate approximately 26 BNS merger events per year within 500 Mpc that are covered by the catalog.

We proceeded by assigning 1170 random timepoints to represent BNS merger events, corresponding to a simulated period of 45 years. These events are uniformly distributed between $\mathrm{MJD}=60087$ and $\mathrm{MJD}=60835$. Since we assume that all GWEM events are distributed with time uniformly and randomly, we divide these events randomly into 28 parts following Poisson distribution. This random assignment of timepoints effectively simulates the equivalent of 28 full O4 observation cycles in a Monte Carlo-like method. 

Another critical parameter in the simulation is the kilonova intrinsic opacity $\kappa$, which significantly impacts the kilonova luminosity but remains highly uncertain. As a preliminary simulation, we consider here only the case where the extinction of the projectile is independent of the inclination, i.e., for a kilonova, the intrinsic extinction is a deterministic value. Theoretical studies, like Metzger et al. (2010), suggest that the opacity is linked to the column density of lanthanide elements in the BNS merger ejecta, which is represented in units of $\mathrm{cm^{2}\ g}^{-1}$. Estimates of $\kappa$ for BNS mergers ranging from as high as $>20$ (\cite{Tanaka2020}) to $\sim0.2$ (\cite{Coughlin2019}) using AT2017gfo, sole kilonova of the BNS merger event GW170817.

\begin{figure}
    \centering
    \includegraphics[width=9cm]{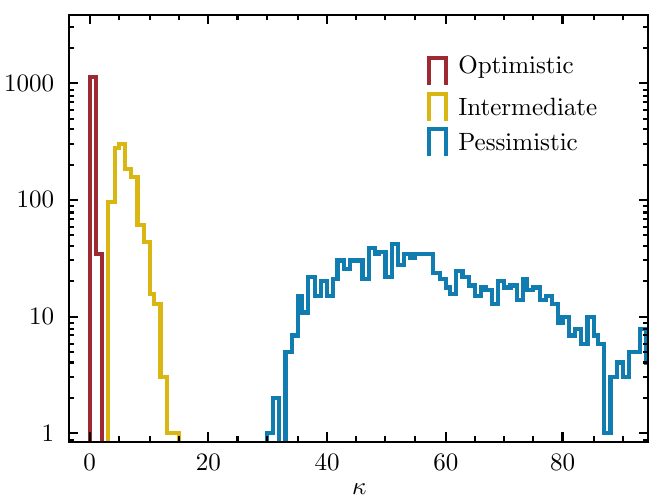}
    \caption{Different presets of opacity $\kappa$ applied in our simulation.}
    \label{fig:kappa}
\end{figure}

To accommodate these discrepancies, we established three distinct opacity intervals: an optimistic range of 0.2-1.0, an intermediate range of 2-10, and a pessimistic range of 20-100, following modified $\Gamma$ distribution as $\Gamma_{\alpha=3,\beta}(\kappa-\kappa_\mathrm{min})$ where $\kappa_\mathrm{min}$ is the assumed minimum value of $\kappa$. These distributions are shown in Figure \ref{fig:kappa}. As the $\kappa$ increases, there is a marked decrease in luminosity and a faster decline in the early kilonova stages, rendering BNS mergers more challenging to be detected under the intermediate and pessimistic cases.

Focusing on the attributes of single neutron stars, in this simulation we utilize the ARP4 equation of state (EOS) model (\cite{PhysRevD.88.044026}), which enables us to derive the neutron star's parameters (such as baryonic mass $M_b$ and compactness $C$ from its gravitational mass $M_g$ through simple polynomial relationships. Specifically, $M_b = 1.33M_g - 0.29$ and tidal deformability $\Lambda_D = 6000M_{\rm g}^{-5}$ is also calculated accordingly. Therefore, in our simulations, the essence of single neutron stars is primarily dictated by their gravitational mass.

Landry and Read (2021) noted that NSs in gravitational wave BNS systems, including GW170817, appear to have elevated masses. Therefore, we assume that the NS masses in the GW BNS systems are not uniform, with one having a gravitational mass of $M_{g,1}=1.1\pm0.3 M_\odot$, the component of the NS mass measured in a relatively wide-separated BNS (e.g., \cite{_zel_2016}), and the other having a significantly large coverage of gravitational mass of $M_{g,2}=1.4^{+ 1.3}_{-0.3} M_\odot$. The reason for heavier $M_{g,2}$ is that extragalactic BNS merger events are often observed to have a heavier NS mass (e.g., \cite{Landry2021}), thus M2 is chosen here to have a slightly higher NS mass distribution than $M_{g,1}$, while the setup of the different companion mass distributions in the BNS system is similar to previous works (e.g., \cite{Farrow2019}, \cite{Frostig2022}).

Through this methodology, all necessary simulation parameters are derived from our predefined settings, thereby enabling us to explore the optical characteristics of BNS mergers and their associated GWEMs, specifically kilonova, under the various parameter configurations we have set.

\subsection{Generating multi-messenger signals}

The detection of the electromagnetic counterparts of gravitational waves requires the observation and analysis of gravitational wave signals. A key aspect of this process is the fact that different ground-based gravitational wave detectors will produce separate signals, resulting in different spatial probability distributions of the GW source on the celestial sphere-hereafter referred to as GW maps. 

By combining the signals from multiple arrays, the ``true'' location of the gravitational wave source should manifest itself as an enhanced GW map value. The refined GW map will then be distributed through the GRACE network to various observing facilities. This allows optical telescopes, including {\it SiTian}, to plan their observations by checking the GW map.

\subsubsection{Simulating GW signals}

The generation of the GW map is what we discuss first. A GW map can be derived from GW waveform data observed by ground-based GW detectors, therefore it is important to determine the active detectors for GW map generation. During the O4 phase, there are four active ground-based GW arrays active and capable of capturing BNS merger events: the LIGO Hanford / Livingston (abbreviated as H and L), Virgo (V), and KAGRA. However, due to KAGRA's limited operational hours, restricted sensitivity, and potential disruption from an earthquake in early 2024, we exclude KAGRA's contribution in our simulations.

During the O4a phase, which lasting approximately eight months, only H and L were operating, with uptime rates of about 69\% and 71\% in total, respectively. Following a three-month maintenance and upgrade period, O4b was then set to operate for 14 months. Both H and L continued their observations during O4b (assumed with the same uptime rates as O4a). Meanwhile, Virgo (\cite{Acernese2014}) began operating in O4b and was assumed to be active throughout the entire O4b phase. This configuration allows for joint HLV observations during O4b, enhancing the reliability of the GW map, as we will discuss in section 3.

In Section 2.1, we uniformly and randomly assigned 1170 BNS merger events to occur during O4. Considering our assumptions about the operation times of the HLV detectors above, we deduce that the likelihood that a BNS merger happening in O4a and O4b is approximately 40\% and 48\%, respectively. Under these circumstances, a BNS merger could be detected by 0 to 2 (non-detection, either H / L, or both H+L) in O4a or 1 to 3 detectors (H+V, L+V, or H+L+V) in O4b. Still, there is also a 12\% chance that a BNS event will occur during the maintenance period, resulting to the BNS merger event undetectable.

\begin{figure}
    \centering
    \includegraphics[width=14cm]{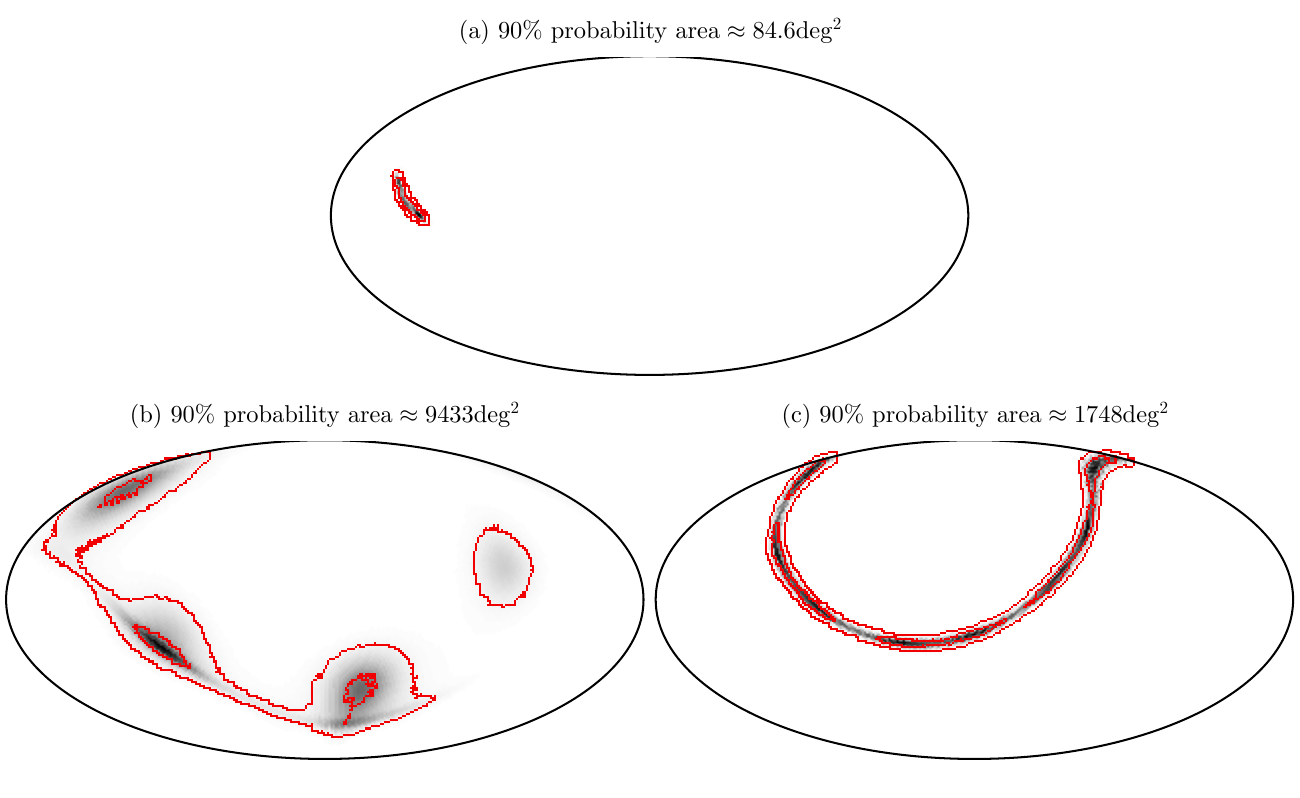}
    \caption{Example GW maps generated during our simulation, showing (a) single peak, as a resulted GW map of two GW detectors, H+L in top panel; (b) multiple peaks, with a single GW detector in bottom-left panel. Four peaks can be seen in this GW map; and (c) unresolved multi-peaks, with a single GW detector in bottom-right panel.}
    \label{fig:gw_map}
\end{figure}

Once the detectors within the HLV network that have detected a BNS merger event, a GW map for that event can be generated using the {\tt bilby} package. With the input of relevant parameters and detector data, {\tt bilby} produces a GW map that adheres to the {\tt HEALPix} format, as demonstrated in Figure \ref{fig:gw_map}.

Moreover, {\tt bilby} provides the GW signal-to-noise ratio (GW SNR) for each detector, allowing us to assess whether a GW signal can be detected based on SNR levels. In our simulation, we classify BNS merger events into three categories according to their combined GW SNR: Class I (combined $\mathrm{GW\ SNR} > 20$), Class II ($9 < \mathrm{GW\ SNR} < 20$), and Class III ($\mathrm{GW\ SNR} < 9$, indicating undetected events).

Every BNS merger event is thus associated with its unique GW map along with the corresponding GW SNR, which indicates whether the event was detected by the HLV network, the strength of its GW signal, and the specific spatial distribution of the GW source on the GW map.

\begin{figure}
    \centering
    \includegraphics[width=14cm]{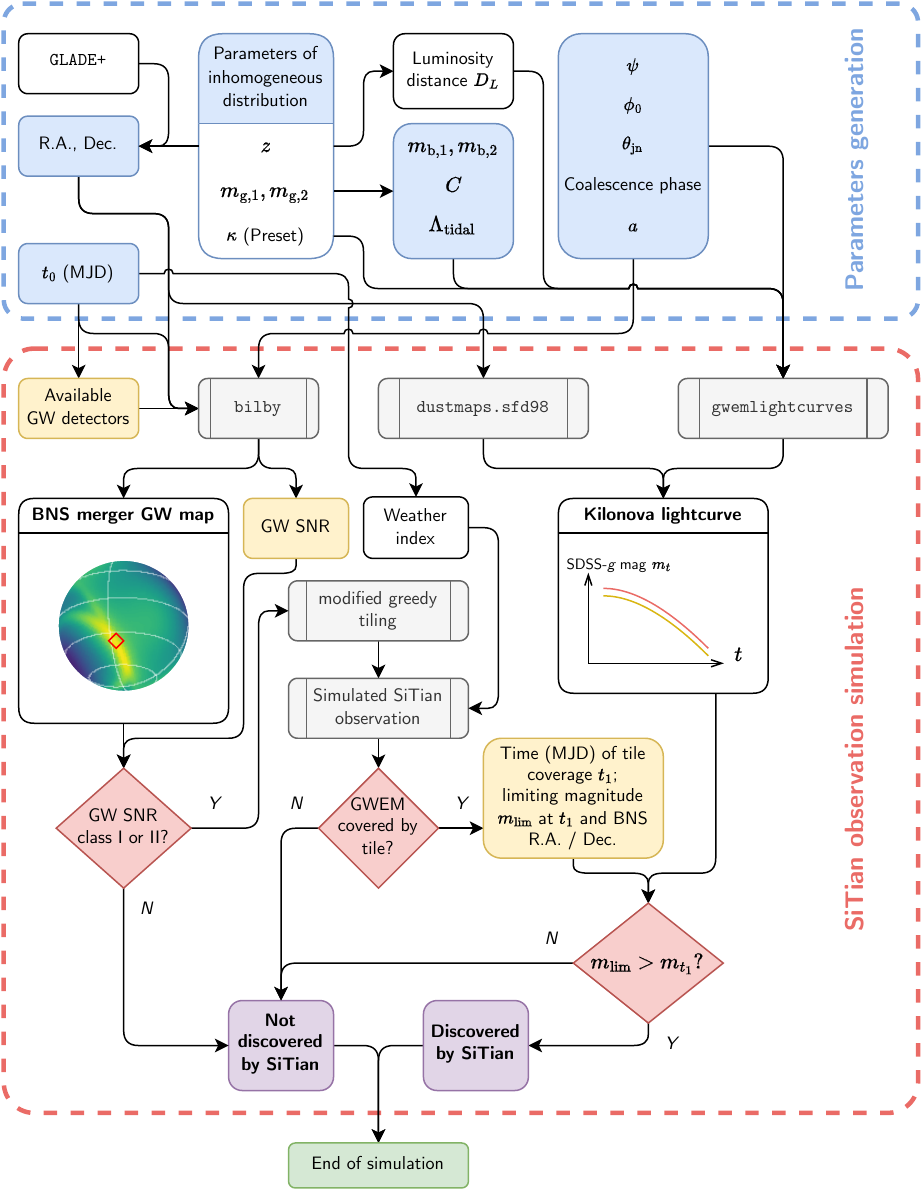}
    \caption{Flowchart illustrating our full simulation.}
    \label{fig:flow}
\end{figure}

\subsubsection{Simulating {\it SiTian} Follow-up Observations for GWEMs}

{\it SiTian}'s follow-up observations for BNS mergers can be broadly divided into three stages: tiling of telescope FoV (``tile'' in the following contents) based on the GW map which is termed ``tiling'', identifying and prioritizing the most promising tiles for GWEM search, and finally determining whether the GWEM is bright enough to be detected by {\it SiTian}, i.e., whether a kilonova will be discovered.

\begin{enumerate}

\item {\it Tiling Process}

A natural idea for observation planning using GW maps is to spread the tiles of the telescope over the observable area using certain methods, and to search for those tiles that have a high spatial distribution probability values. Following this idea, we do the following for each GW map: interpolate the GW map to higher resolution, and find the coordinate of the point with the highest probability value in the interpolated GW map. We use this coordinate as a reference point, place a {\it SiTian} tile at this center, and gradually arrange the tiles along the right ascension and declination directions with a certain overlap (default 10\%) until the observable area of {\it SiTian}, i.e. area with $\mathrm{dec}>-19^\circ$, is filled with tiles. Still, due to the temporary unsolvable problem of FoV distortion near the celestial pole, we did not perform tiling for the region with a \textit{declination} $>85^\circ$. For each tile, all interpolated pixel points located within the given tile were extracted, and the mean of probability values of these pixel points was computed and used as the probability value for the given tile.

When interpolating, we use the {\tt Healpy} program to interpolate the GW map to $\mathrm{NSide=128}$. This is based on two considerations: on the one hand, generating a tile for a given GW map requires the coordinate of the point with the highest probability value. The average pixel distance for $\mathrm{NSide=128}$ is about $0.46^\circ$, far below the {\it SiTian}'s tile edge length of $5^\circ$. The coordinate of the point with the highest probability value given by each GW map is thus precise enough for tiling. On the other hand, all pixels contained in a tile are to be extracted when calculating the probability value for each tile. The setup of $\mathrm{NSide=128}$ allows a coverage more than 100 pixels in each tile, ensuring a fast processing speed while maintaining a satisfactory accuracy when calculating the probability value.

Previous steps give us the tiles generated for {\it SiTian} observations from the GW map, as well as the probability value for each tile, which allows us to further schedule the simulated observations.

%Each detected BNS merger event comes with a corresponding GW map. Our tiling approach employs a dynamic method that centers on the celestial coordinates where the GW map reaches its peak value. Using {\tt healpy}'s interpolation algorithm, we interpolate a single GW map to $\mathrm{NSide}=128$ and pinpoint the celestial coordinates of maximum probability. $\mathrm{NSide}=128$ has a mean pixel spacing of $0.46^\circ$, which is small enough below the $5^\circ$ edge length of a single SiTian tile, allowing us to obtain acceptable results at less cost when calculating the probability value of the spatial distribution covered by each SiTian tile. This helps us identify the spot where {\it SiTian}'s field of view (FoV) intersects with the highest probability on the GW map. The FoV is used to tiling the entire observable sky area on the celestial sphere, centered on this point. Given the Xinglong Observatory's latitude, observable sky regions are those with declinations between $+85^\circ$ and $-19^\circ$. Due to FoV aberration issues at high declinations, sky regions with declination $>85^\circ$ are not considered in this simulation. Tiles overlap by 10\% by default, typically resulting in about 1500 tiles.

\item {\it Sorting Tiles by GW Map}

A simple greedy algorithm can be used to give a reasonable tiling sequence by arranging tiles by probability value calculated previously in ascending order, then prioritizing those with higher GW source probabilities. However, for GW sources detected by 1 or 2 GW detectors, the GW maps may present multiple peaks (as illustrated in Fig. \ref{fig:gw_map}(b) and (c)), which could cause a simple greedy algorithm to mix up tiles with similar GW map values across the celestial sphere.

We thus introduce a modified greedy algorithm. After initially sorting tiles, we select the first 50 tiles, convert their center coordinates into a Haversine distance matrix, and apply the {\tt DBSCAN} clustering method from the {\tt scipy} package. Given that GW maps from 1-2 GW detectors often show 2-4 probability peaks, if {\tt DBSCAN} yields 2-4 clusters, the modified greedy algorithm will extract the largest GW map value from each cluster and reorder these clusters accordingly. By doing so, during actual search operations, {\it SiTian} will observe the cluster with the highest probability first, followed by the next cluster of the most probable, and so forth, ensuring a more efficient search strategy for optical time-domain survey.

\item {\it Assessing Discoverability of the Kilonova During Observation}

While we can efficiently generate a ranked tiling list, directly applying kilonova light curves to test observability is not precise. Several corrections and considerations must be made, including adjustments for Galactic extinction and the various effects impacting {\it SiTian} observations. For instance, the exposure time of {\it SiTian}, assuming the use of the SDSS-{\it g} band with a commercial CMOS camera, we estimate a 3-minute, 5-$\sigma$ limiting magnitude of 21.5 mag, a night sky background of $21.0\ \mathrm{mag\ arcsec^{-2}}$, a seeing of $2''$, and a system efficiency of 40\%. For adequate depth, we assume a search exposure time of 5 minutes, resulting in a limiting magnitude of 22.0 mag for SDSS-{\it g} (not considering multiple exposures or coordinated observations by multiple {\it SiTian} arrays).
\end{enumerate}

Additionally, {\it SiTian} faces typical challenges associated with ground-based optical surveys. Environmental factors are categorized into four groups: sky brightness, atmospheric extinction, moonlight, and weather conditions. At the Xinglong Observatory, sky brightness mainly arises from night sky and the Sun during morning and evening twilights. Night sky brightness is accounted for in the 22.0 mag limit; thus, we concentrate on sunlight near morning twilight. Assuming the Sun is not visible when it's less than $-6$ degrees below the horizon, the reduction in limiting magnitude $\Delta m_\odot$ at zenith due to Sun's altitude $\mathrm{alt_\odot}$ (between $-6^\circ$ and $-16^\circ$) follows:
$$\Delta m_\odot = 10.6695 + 1.96849 \cdot \text{alt}_\odot + 0.121029 \cdot \text{alt}_\odot^2 + 0.00247973 \cdot \text{alt}_\odot^3$$

Next, assuming atmospheric extinction is determined solely by airmass, we calculate airmass using {\tt airmassSpherical} function installed in {\tt PyAstronomy}. Experience shows that the relationship between atmospheric extinction in the SDSS-{\it g} band at Xinglong Observatory and airmass is expressed as $\Delta m_\text{airmass} = 0.60 \cdot \text{airmass}$.\footnote{Our two empirical formulas here were obtained from our December 2023 {\it in situ} observations at Xinglong Observatory. We note that in our experience, atmospheric extinction here can be overestimated as a result of poor meteorological conditions during the day or rapid deterioration of the seeing after nightfall.}

Moonlight effects are estimated empirically, which depend on both lunar phase and distance, so we approximate a linear decrease in limiting magnitude for lunar distances $<30^\circ$, a decrease of 3 magnitudes for a full Moon at a 1-degree distance, and targets closer than 0.3 degrees to the Moon as obscured and unobservable. Lunar position and phase are computed using the {\tt PyEphem} package.

Finally, considering these environmental variables, we adjust the expected kilonova brightness to evaluate whether it exceeds {\it SiTian}'s adjusted limiting magnitude, thereby determining if the kilonova is discoverable during the simulated observations.

Weather is one of the most intricate and impactful aspects of ground-based astronomical observations. Due to the seasonal variations in weather patterns at the Xinglong Observatory influenced by monsoons, detailed weather modeling is outside the scope of this simulation. Instead, we empolyed historical meteorological data from two sources: the European Centre for Medium-Range Weather Forecasts (ECMWF) CAMS global reanalysis ({\tt EAC4}) and internet archive of historical meteorological records between January 1, 2019, 00:00 hours to December 31, 2020, 22:00 hours at the Xinglong Observatory's location, to represent the actual weather conditions.

To account for weather effects, we introduce weather index (WI) to determine the probability of exposure success. The value of WI varies between 0 and 1 and inversely relates to the exposure success rate $p=1-WI(t)$. Conditions like high humidity, precipitation, and strong winds will render direct observation impossible, which would assigning a WI value of 1 (unfavorable). Ideal weather with a surface wind speed below $3\ \mathrm{m\ s^{-1}}$ equates to a WI of 0. Intermediate conditions result in a WI that is a linear combination of the differences in wind speed, atmospheric transparency, and normalized cloud cover.

Based on the WI, there are about 219 average observable nights per year, which aligns with the actual annual average of around 200 observable nights at Xinglong Observatory, validating the use of WI to represent weather conditions.

The process of determining whether {\it SiTian} can detect a kilonova is summarized as follows: At time $t_0$, {\tt GRACE} distributes the data to the GW map. After about 1 minute of downloading and tiling, the telescope status is checked. If conditions are safe for observation and the target tile is above 20 degrees above the horizon, the kilonova search will begins with a 5-minute preparation time (we anticipate the terminate of ongoing observation and immediate initialization of GWEM search in actual observation scenario). During the GWEM searching, each tiling takes given exposure time (5 minute by default, or 15 / 30 minutes, see section below) plus 1 minute of telescope slewing time, and the success of the exposure is randomly determined based on the WI value determined by simulated time. Failed exposures are repeated until successful.

The system constantly monitors whether the kilonova is within the current tiling range. When the kilonova is found within the tiling and a successful exposure occurs, the loop ends. The kilonova's SDSS-{\it g} band magnitude is obtained using {\tt gwemlightcurves}, along with the {\tt SFD98} (\cite{sfd98}) $E(B-V)$ extinction value and the reduction in limiting magnitude $\Delta m$ due to various factors. If the magnitude of the kilonova with extinction ($m_\mathrm{KN}+3.30\times E_\mathrm{SFD98}(B-V)$) is less than the adjusted limiting magnitude ($22.0-\Delta m$), the simulation announces a successful kilonova discovery and records the moment of end (in MJD), the number of tiles used, combined GW SNR and SNR classification, and the magnitude difference between the adjusted limiting magnitude and actual magnitude. Note that the time resolution of the {\tt gwemlightcurves} program is only 0.1 days. Considering that the early-postmerger stage lightcurves of a kilonova reasonably have a sharp ascending phase, we used a simple piecewise function model to represent its very early light curves characteristics, i.e., the absolute magnitude is constant at postmerger 0.05-0.1 days to the magnitude at $t = 0.1$ days, while at days 0-0.05 it decreases linearly from magnitude 10.

In conclusion, we have constructed a comprehensive simulation framework that covers the generation of BNS merger events through to the multi-factorial consideration for {\it SiTian}'s optical search for kilonova. The following section discusses the simulation outcomes of 1170 sources and performs an analysis.

\begin{figure}
    \centering
    \includegraphics[width=14cm]{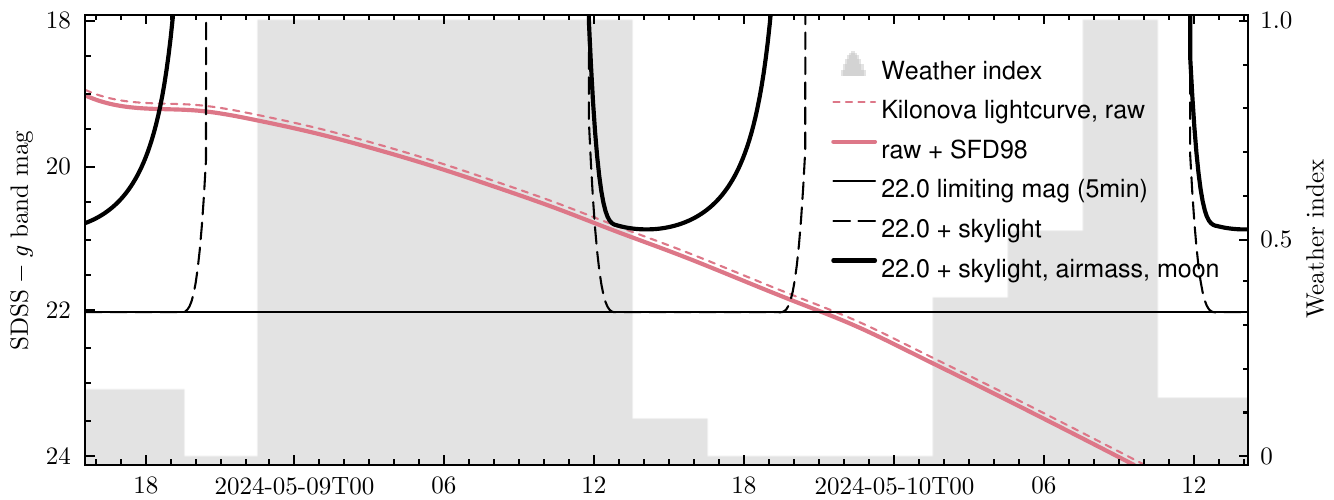}
    \caption{An example of a bright kilonova lightcurve, showing the effect of environmental factors, manifested as elevated limiting magnitude and weather index, making the source not observable for a majority of times.}
    \label{fig:lc_sing}
\end{figure}

\section{Simulation results}

Our simulation derived the characteristics of 1170 BNS merger sources, generating their GW profiles (including the number of active detectors, GW map, and GW SNR) and corresponding kilonova SDSS-{\it g} band light curves with 3 different levels of intrinsic opacity ($\kappa$). A modified greedy algorithm, informed by actual observational experience, \textit{is employed to simulate {\it SiTian}'s kilonova search during the O4 period.}

Firstly, we examine the GW observational properties of the BNS sources. As detailed in Section 2.2.1, these BNS merger events were classified into three categories based on GW SNR: Class I (combined $\mathrm{GW\ SNR} > 20$), Class II (combined $9 < \mathrm{GW\ SNR} < 20$), and Class III (combined $\mathrm{GW\ SNR} < 9$). Among the 1170 simulated BNS mergers, 17 fall into Class I, indicating that the HLV network is expected to detect $0.29\pm0.54$ Class I sources during a single O4 period. There are also 163 Class II sources, translating to an O4 average of $3.4\pm1.8$ detections. Therefore, a total of 180 BNS mergers are potentially detectable via ground-based GW observations, averaging $3.7\pm1.9$ cases per O4 period. These details are presented in Table 2.

\begin{table}[h]
\centering
\caption{GW BNS classification summary}
\label{tab:BNS_classification}
\begin{tabular}{@{}lccccc@{}}
\toprule
\textbf{GW BNS classification} & \textbf{Count} & \multicolumn{3}{c}{\textbf{Number of detectors active}} & \textbf{O4 average} \\ \cmidrule(lr){3-5}
 &  & \textbf{3} & \textbf{2} & \textbf{1} &  \\ \midrule
Class I & 7 & 6 (86\%) & 0 & 1 (14\%) & $0.29\pm0.54$ \\
Class II & 82 & 58 (71\%) & 17 (21\%) & 7 (9\%) & $3.4\pm1.8$ \\
Class III & 378 & 140 (37\%) & 120 (32\%) & 118 (31\%) & - \\
No GW BNS & 703 & - & - & - & - \\ \midrule
Sum & 1170 & 204 (17\%) & 137 (11\%) & 126 (11\%) & $3.7\pm1.9$ (detected by GW detectors) \\ \bottomrule
\end{tabular}
\end{table}

The remaining 539 Class III events and 451 sources without detectable GW signals are excluded from further discussion due to their low GW SNR or non-detections.

\begin{figure}
    \centering
    \includegraphics[width=9cm]{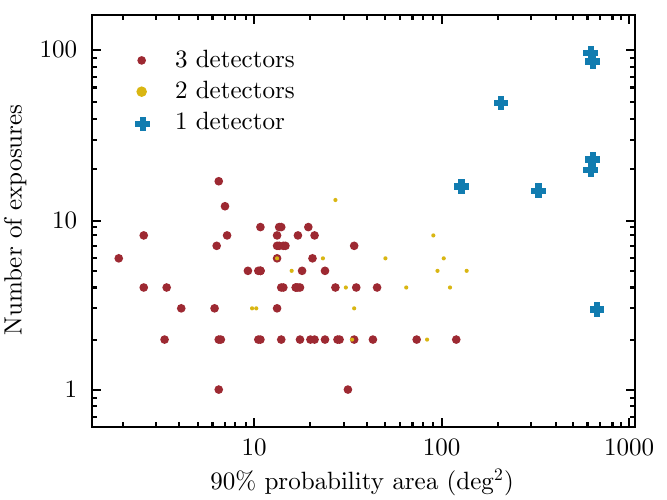}
    \caption{Scatter plot showing the relation between 90\% probability area and number of tiling exposures (5min each case). A good positive relation between 90\% probability area and exposures can be seen in this plot.}
    \label{fig:90scatter}
\end{figure}

\begin{figure}
    \centering
    \includegraphics[width=14cm]{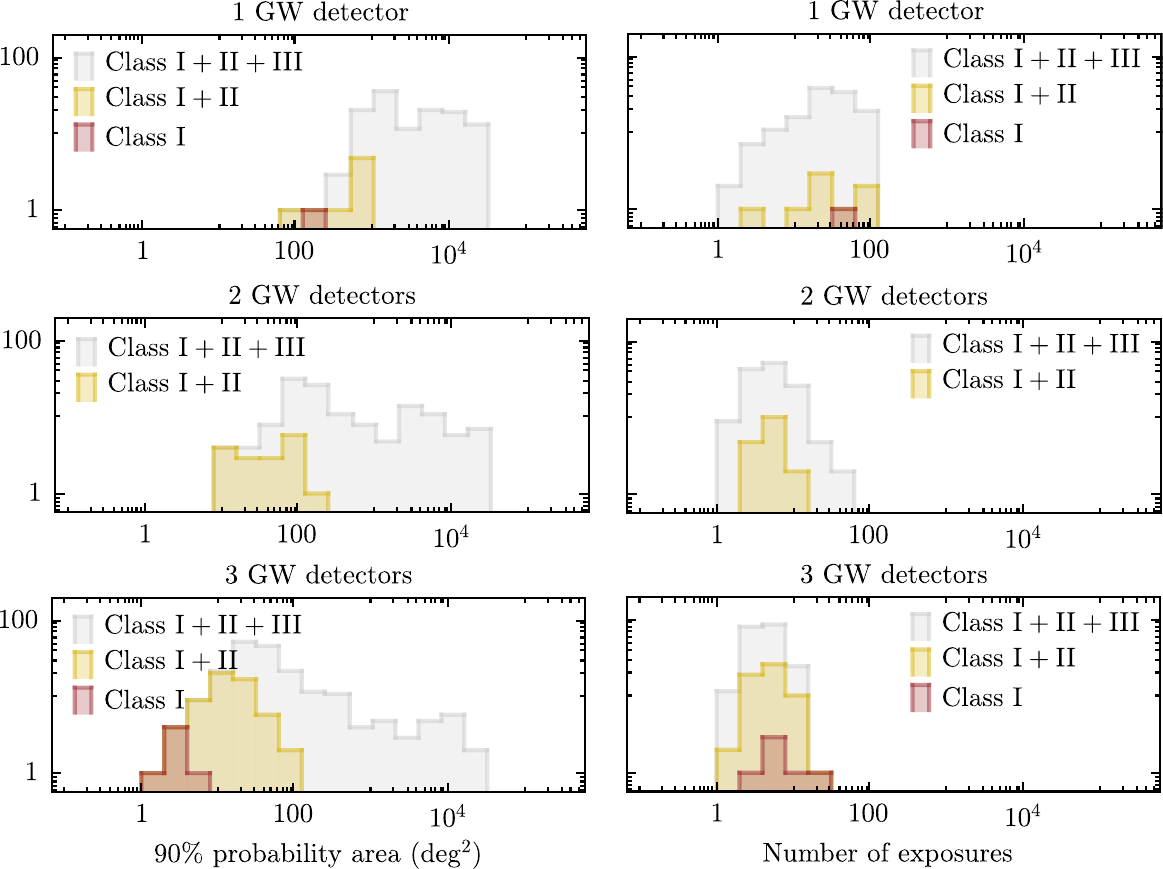}
    \caption{Left: Histogram of 90\% probability area classified by detector counts and GW SNR. Right: same classification but the histogram of tiling exposure numbers, and Class III sources were removed because they can not be observed.}
    \label{fig:histo}
\end{figure}

However, we note that the GW SNR alone does not fully characterize the diversity of GW maps for BNS mergers. As shown in Figure \ref{fig:gw_map}, the different numbers of GW detectors active for certain BNS merger event significantly impacts the characteristics of GW map. Thus, we analyze the nature of the GW map in conjunction with the number of {\it SiTian} tiling exposure counts in the simulation. Since the GW map represents a spatial probability distribution of the BNS merger, we quantify the characteristics of a given GW map using the 90\% probability area (90\% probability area hereafter), which denotes the region where the BNS has a 90\% likelihood of occurrence.

The resulting data has been visualized in Figures \ref{fig:90scatter} and \ref{fig:histo}. Our simulations lead to the following conclusions:

\begin{enumerate}
    \item The number of {\it SiTian} exposures anticipated in the simulations displays a negative correlation with the GW SNR, as well as a positive correlation with the 90\% probability area. Specifically, smaller 90\% probability areas correspond to fewer expected exposures.
    \item The expected number of {\it SiTian} tilings is also related to the number of GW detectors contributing to the detection. When only one detector is involved, simulations generally predict 10 to 80 tilings for {\it SiTian}. However, when 2-3 detectors operate concurrently, the expected number of tilings reduces to approximately 1-20.
    \item The use of multiple GW detectors significantly enhances the GW SNR. Notably, the 90\% probability area of the GW map exhibits a robust negative correlation with the GW SNR, especially when considering the fixed number of detectors detecting a given source. As the 90\% probability area positively correlates with the number of required exposures from simulations, both the combined efforts of multiple GW detectors and high GW SNR favor {\it SiTian}'s search.
    \item Specifically, when three GW arrays operate concurrently, the 90\% probability area can shrink to less than 10 square degrees, smaller than {\it SiTian}'s FoV. This is particularly advantageous for {\it SiTian} searches, though there are still factors such as the narrow distribution of spatial locations, a need up to 10 tilings would confidently cover the spatial region where the BNS merger resides.
\end{enumerate}

\begin{figure}
    \centering
    \includegraphics[width=9cm]{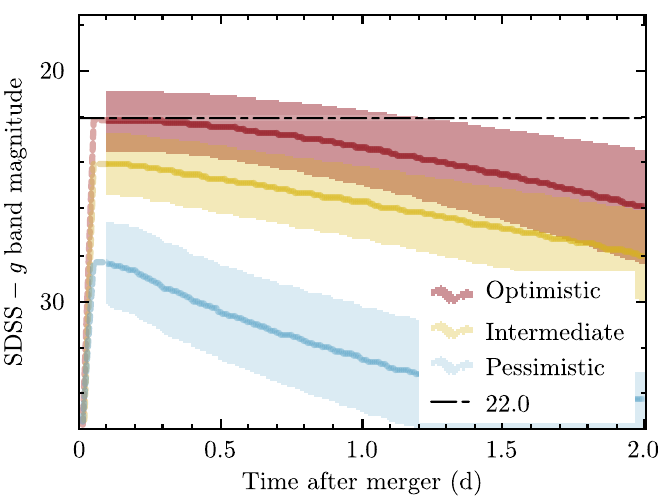}
    \caption{Mean and $2-\sigma$ interval of kilonova SDSS-{\it g} band lightcurve with 3 opacity presets as a function of time. 5-min SDSS-{\it g} band limiting magnitudes of 22.0 is shown in black dashed line (see legend). The 0.1 - day gap postmerger is caused by the time resolution of the {\tt gwemlightcurves} program being only 0.1 days, while at the same time the luminosity of the kilonova decreases rapidly in the SDSS-g band, and therefore the lightcurve in the rising phase of the kilonova is only represented by the simple model we constructed (colored dashed line).}
    \label{fig:my_label}
\end{figure}

We now turn our attention to the simulations of {\it SiTian}'s optical band observations. First, we address the characteristics of the generated kilonova lightcurves. As previously mentioned in Section 2.1, we employed three intrinsic opacity presets for the kilonova: 0.2-1.0 (optimistic case, in $\mathrm{cm^2\ g^{-1}}$), 2-10 (intermediate case), and 20-100 (pessimistic case). The lightcurves resulting from these scenarios can be viewed in Figure \ref{fig:my_label}. It is evident that the kilonova lightcurve in the optimistic case is most favorable, with a considerable number of optimistic case kilonova maintaining SDSS-{\it g} band magnitudes brighter than 22.0 magnitudes up to one day post-outburst. Conversely, the initial magnitudes of kilonova in the intermediate and pessimistic cases plummet to 24 and 28 magnitudes, making them too faint to be observed by {\it SiTian}.

To optimize the detection of kilonova light signals, we explore two observation strategies:

\begin{enumerate}
    \item Efficiency-focused approach: Each tile receives a single 5-minute exposure.
    \item Depth-oriented approach: Considering that brighter kilonova peak luminosities often cluster between 21 and 24 mag in the simulations, and that the 90\% probability area of BNS mergers in multiple-GW-detector Class I/II events can be confined to approximately 100 square degrees—allowing coverage within 20 tilings—we simulate scenarios with cumulative exposure times of 15 and 30 minutes per tile. This raises {\it SiTian}'s SDSS-{\it g} band limiting magnitude from 22.0 mag to 22.7 mag and 23.0 mag, respectively.
\end{enumerate}

We now discuss these strategies:

\begin{enumerate}
    \item Efficiency-focused approach: Simulations yield discouraging outcomes. With a BNS merger rate of 100 per year and using the optimistic kilonova opacity preset (0.2-0.8 $\mathrm{cm^2\ g^{-2}}$) in {\it SiTian} observations with 5-minute exposures, we attain only 5 successful kilonova confirmations, translating to an average of $0.17^{+0.40}_{-0.16}$ detections per O4 run. Under intermediate and pessimistic opacity assumptions, no kilonova are deemed observable.
    \item Depth-oriented approach: Simulation results improve considerably compared to the efficiency-dominated case. While {\it SiTian} also fails to observe any kilonova event in the intermediate and pessimistic cases, the number of detectable kilonova in the optimistic case rises to 11 (15 min exposure) and 10 (30 min exposure), corresponding to averages of $0.25^{+0.38}_{-0.16}$ and $0.21^{+0.35}_{-0.13}$ cases during O4, respectively. This is a relatively promising result, suggesting that extending exposure times is effective in kilonova searches despite their rapid luminosity decline. Moreover, if {\it SiTian} operates continuously during the O4 period, there is an optimistic probability exceeding 1/3 that it can confirm at least one kilonova of a GW-detected BNS merger event.
\end{enumerate}

The summary of these findings is presented in Table 3.

\begin{table}
\centering
\caption{Opacity pre-set classification summary}
\label{tab:opacity_classification}
\begin{tabular}{@{}lcccccc@{}}
\toprule
\multirow{2}{*}{\textbf{$\kappa$ presets}} & \multicolumn{3}{c|}{\textbf{Exposure time (Class I+II)}} & \multicolumn{3}{c|}{\textbf{O4 average}} \\ 
\cmidrule(lr){2-4} \cmidrule(lr){5-7}
 & \textbf{5 min} & \textbf{15 min} & \textbf{30 min} & \textbf{5 min} & \textbf{15 min} & \textbf{30 min} \\ \midrule
Pessimistic & 0 (0+0) & 0 (0+0) & 0 (0+0) & 0 & 0 & 0 \\
Intermediate & 0 (0+0) & 0 (0+0) & 0 (0+0) & 0 & 0 & 0 \\
Optimistic & 4 (2+2) & 6 (4+2) & 5 (3+2) & 0.17 & 0.25 & 0.21 \\ \bottomrule
\end{tabular}
\end{table}

\section{Summary and Discussion}

This study has modeled the equivalent of 24 O4 runs of BNS merger events observable by the LIGO Hanford, LIGO Livingston, and Virgo (HLV) gravitational wave (GW) detectors. It also investigated the search for kilonovae using the optical time-domain survey instrument, the {\it SiTian} prototype telescope. Our main conclusions are as follows:
\begin{enumerate}
    \item Assuming a BNS merger event rate of 100 cases per cubic Gpc per year, the HLV array is projected to detect $3.7\pm1.9$ observable BNS mergers during O4. Considering our simulations utilize the {\tt GLADE+} catalog, the actual event rate (simulated events plus those likely present in galaxies not cataloged in {\tt GLADE+}) should be approximately 5-7 cases.
    \item Given the substantial uncertainty surrounding the intrinsic opacity of kilonova, we employed three opacity presets. The kilonova luminosity was found to be highly sensitive to opacity, with only the most optimistic preset yielding detectable kilonova events in our simulations.
    \item Despite kilonova's rapid luminosity decay, extending {\it SiTian}'s exposure time can significantly enhance the detection of fainter kilonova. In the optimistic scenario, increasing the exposure time from 5 to 15 minutes boosts the number of kilonova detected by {\it SiTian} from 0.17 to 0.25 (See Table 3). This also implies that with appropriate exposure times, {\it SiTian} stands a 1/4 chance of capturing at least one BNS merger GW counterpart during O4.
\end{enumerate}

Additionally, we like to discuss other factors in our simulation. Weather as a critical determinant of ground-based optical observations, exerts a substantial impact on kilonova searches. In a separate simulation, approximately 1/4 of BNS mergers lose the opportunity for {\it SiTian} coverage due to unfavorable weather conditions. Although, as mentioned in Section 2, weather is governed by the probabilistic weather index, causing variations among our simulations, we refrain from providing an exact figure. Instead, we semi-quantitatively propose that observable kilonova could increase by roughly 1/3 under perfectly ideal weather conditions, advocating for conducting observations at locations with superior weather conditions, such as the Lenghu Observatory.

The sensitivity of BNS merger detections will be enhanced through technical upgrades to the HLV array. By O5, the maximum distance for BNS merger events detectable by H/L with GW SNR = 9 is expected to extend to ~300 Mpc, and the sensitivity of the KAGRA detector will also improve. Our simulations did not account for changes in HLV detection sensitivity, hence we lack confidence in directly applying our current simulation results to HLV + {\it SiTian} observations during O5.

Similarly, we did not consider the scenario of future joint observations involving multiple {\it SiTian} telescopes. With a fully operational {\it SiTian} node consisting of 18 {\it SiTian} three-color arrays, the exposure time could be reduced to 1/18 of its current value reaching desired limiting magnitude. Furthermore, Class I + II BNS events simultaneously detected by 2-3 GW detectors (as seen in Fig. \ref{fig:gw_map}) could be covered within 20 first tiles. This would enable {\it SiTian} to simultaneously image a considerable portion of the sky, making it highly likely that a single exposure would encompass the spatial location of the BNS, enabling the detection of fainter kilonova with longer exposures. Given that extended exposures with a single telescope can boost kilonova discovery rates, we anticipate that observations with the {\it SiTian} array will significantly enhance kilonova discoveries.

More detailed simulations will be performed for the upcoming O5 with the first {\it SiTian} three-color node.

{\it Data availability}: Processing file data may be provided kindly for reasonable requests. However, since all of the data in this study was randomly generated, we do not provide structured catalog data here.

{\it Softwares used in this work}: {\tt numpy} (\cite{numpy}), {\tt scipy} (\cite{scipy}), {\tt topcat} (\cite{topcat}), {\tt gwemlightcurves} (\cite{Coughlin2018}), {\tt bilby} (\cite{Ashton2019}), {\tt Healpy} (\cite{healpy}), {\tt PyAstronomy} (\cite{pya}), {\tt astropy} (\cite{astropy}), {\tt PyEphem} (\cite{ephem}), {\tt dustmaps} (\cite{dustmaps})

\section*{Acknowledgements}

The SiTian project is a next-generation, large-scale time-domain survey designed to build an array of over 60 optical telescopes, primarily located at observatory sites in China. This array will enable single-exposure observations of the entire northern hemisphere night sky with a cadence of only 30-minute, capturing true color (gri) time-series data down to about 21 mag. This project is proposed and led by the National Astronomical Observatories, Chinese Academy of Sciences (NAOC). As the pathfinder for the SiTian project, the Mini-SiTian project utilizes an array of three 30 cm telescopes to simulate a single node of the full SiTian array. The Mini-SiTian has begun its survey since November 2022. The SiTian and Mini-SiTian have been supported from the Strategic Pioneer Program of the Astronomy Large-Scale Scientific Facility, Chinese Academy of Sciences and the Science and Education Integration Funding of University of Chinese Academy of Sciences.

We thank the reviewers for a series of constructive comments on this paper.

This work is supported by the Strategic Priority Research Program of the Chinese Academy of Sciences, Grant No. XDB0550103.
Y.H. and J.F.L. acknowledges the National Key RD Program of China Nos. 2023YFA1608300.
Y.H. acknowledges the supported from the National Science Foundation of China (NSFC Nos. of 12422303 and 12261141690).
J.F.L. acknowledges support the NSFC through grant Nos. of 11988101 and 11933004, and support from the New Cornerstone Science Foundation through the New Cornerstone Investigator Program and the XPLORER PRIZE.

\bibliographystyle{raa}
\bibliography{ms2024-0354}

%
%               one-column-spanning table
%________________________________________ Table 1: Use_of_the routines

\label{lastpage}

\end{document}